\documentclass[letterpaper, 10 pt, conference]{ieeeconf}  

\IEEEoverridecommandlockouts                              

\usepackage{comment}
\usepackage{soul}
\usepackage{amsmath}
\usepackage{amsthm}
\usepackage{textcomp}
\usepackage{nicefrac}
\usepackage{amsfonts}
\usepackage{algorithm}
\usepackage{algorithmic}
\usepackage{multirow}
\usepackage{colortbl}
\usepackage{color}
\usepackage[table]{xcolor}
\usepackage{epigraph}
\usepackage{caption}
\usepackage{subcaption}
\usepackage{hyperref}
\usepackage{tabularx}
\usepackage{float}
\usepackage{longtable}
\usepackage[pdftex]{graphicx}

 \newtheorem{Theorem}{Theorem}
 \newtheorem{Lemma}{Lemma}

 \newtheorem{Proposition}[Theorem]{Proposition}
 \newtheorem{Remark} {Remark}
 \newtheorem{Example} {Example}

\newcommand{\R}{\mathbb R}

\newcommand{\diag}{\mathop{\mathrm{diag}}}

\providecommand{\keywords}[1]{\textbf{\textit{Index terms---}} #1}

\newcommand{\RT}[1]{ {\color{brown} (RT:{#1}) } }
\newcommand{\RM}[1]{ {\color{magenta} (RM:{#1}) } }

\usepackage{calc}
\newsavebox\CBox
\newcommand\hcancel[2][0.5pt]{%
  \ifmmode\sbox\CBox{$#2$}\else\sbox\CBox{#2}\fi%
  \makebox[0pt][l]{\usebox\CBox}%
  \rule[0.5\ht\CBox-#1/2]{\wd\CBox}{#1}}

\DeclareMathOperator*{\argmin}{argmin}

\newcommand*\diff{\mathop{}\!\mathrm{d}}

\usepackage{algorithmic}
\usepackage{algorithm}

\title{\LARGE \bf
Analysis of the Identifying Regulation\\with Adversarial Surrogates Algorithm}

\author{Ron Teichner$^{1}$, Ron Meir$^{1}$ and Michael Margaliot$^{2}$
\thanks{$^{1}$RT and RM are with the Viterbi Dept. of Elec. \& Computer Eng., Technion, Israel Institute of Technology, 
        {\tt\small ron.teichner@gmail.com}. They are partially supported by ISF grant 1693/22 and by the Skillman chair (RM).}%
\thanks{$^{2}$MM  is
  the incumbent of the Systems and Control Theory Chair   at  the School of Elec.  Eng.,
  Tel-Aviv University, Israel 6997801.}%
}

\begin{document}

\maketitle
\thispagestyle{empty}
\pagestyle{empty}
\ 

\begin{abstract}
Given a  time-series of {noisy} 
measured outputs of a dynamical system~$\left\{z^k\right\}_{k=1}^N$,     the Identifying Regulation with Adversarial Surrogates~(IRAS) algorithm aims to find a non-trivial first integral of the system, namely, a scalar function~$g$ such that $ g(z^i) \approx g(z^j)$, for all~$i,j$.
IRAS has been suggested recently and was used successfully 
in several learning tasks in models from biology and physics. 
Here, we give the first rigorous analysis of this algorithm  in a specific  setting. We assume that the observations  admit a linear first integral and that they are contaminated  by  Gaussian noise. We show that in this case the IRAS iterations are closely related to the self-consistent-field~(SCF) iterations for solving  a generalized Rayleigh quotient minimization problem. Using this approach, we derive several  sufficient conditions guaranteeing local convergence of IRAS to the linear  
first integral.
\end{abstract}


\section{INTRODUCTION}
A conserved quantity of a dynamical
system is a
quantity that can be expressed using the   time-varying   variables of the system, but remains constant along each
trajectory of the system. A classical example is the {Hamiltonian}
in a mechanical system~\cite{class_mech_golds}.

An important challenge in systems and control theory is to discover conserved quantities in dynamical systems in a 
\emph{data-driven} way, i.e. based on 
  observed trajectory data, and assuming that the 
  underlying dynamical equations are unknown. Machine learning~(ML) algorithms 
  for addressing this problem 
  have found  applications in several important domains of science. In physics, they are used to find symmetries and  conservation laws~\cite{liu2021machine,mototake2021interpretable,greydanus2019hamiltonian,DIERKES2021210}, for example, by
 parameterizing     possible
 Hamiltonians using a learning 
neural network~\cite{greydanus2019hamiltonian}. It is important to note that even if the dynamical equations of a system are known, the problem of analytically determining conserved quantities is non-trivial.

Determining conserved quantities is also important in biological systems~\cite{KAKIZOE201539}.
Classical models in population dynamics
(e.g., the Lotka-Volterra model)
and in mathematical epidemiology  (e.g., the Kermack and McKendrick model) admit a conserved quantity  
(or  a first integral) and this has important biological implications (see, e.g.~\cite{KAKIZOE201539}). 
Living systems maintain stability against internal and external perturbations, a phenomenon
known as homeostasis~\cite{billman2020homeostasis}.
In particular, closed-loop control schemes tightly regulate
  \emph{internal} conserved quantities, and explicitly determining these quantities    is an   important and non-trivial problem~\cite{samad2021}.

The underlying assumption behind  ML algorithms for the discovery of conserved quantities is that there exists a 
function $g:\mathbb{R}^{n} \rightarrow \mathbb{R}$ of the observed variables   {that is conserved along trajectories (but may have a different constant value along different trajectories)}. Note
that optimizing for conservation alone can (and does) lead to trivial quantities such as predicting a constant function~$g(z)\equiv c$,  independently of~$z$.

The difficulties in deriving data-driven 
algorithms for detecting conserved quantities include: 
(1)~the learning is based on a finite data set; (2)~the measured data is typically noisy, thus violating any exact conservation law;   
(3)~the conserved quantities may vary in time, for example,
in  a  physical system drifting  over time to higher or lower energy states;
and  (4)~the algorithm  may output  a  trivial  result such as a function~$g(z)\equiv c$, independent of~$z$. Alet {\it et al.}~\cite{alet2021noether} refer to a non-trivial $g(\cdot)$ as a   \textit{useful} conservation law.

Often, the presentation of such algorithms is accompanied by demonstrating their empirical  success in various test cases, but with no rigorous analysis
of convergence,
 robustness to noise, etc.

 IRAS~\cite{IRAS_PNAS} is an iterative 
data-driven  algorithm for detecting
conserved quantities. In each iteration, it solves 
  a min-max optimization problem.  IRAS was used successfully~\cite{IRAS_PNAS} in the discovery of the conserved
  quantity in several examples including:  a kinetic model of protein 
interactions from~\cite{samad2021}; a psycho-physical experiment in which a PID controller regulated a  stimulus 
signal (see the description in~\cite{marom2011relational}); a model of interactions in an ecological system from~\cite{Murray2002};  and a   Hamiltonian  system from physics.
  Yet,  theoretical analysis of IRAS is lacking.

Here, we analyze 
IRAS in a specific  probabilistic setting,  assuming a linear first integral and Gaussian noise. 
{Linear first integrals are common in closed compartmental systems, like epidemiological  models assuming a constant population size~\cite{SEIR} and 
gross-substitute systems in economics~\cite{Naka79}.  We demonstrate  our theoretical results using an important 
model from systems biology called the  ribosome flow model on a ring~(RFMR)~\cite{RFMR}. }

We provide {explicit}
sufficient  conditions for IRAS convergence  that depend on the signal to noise ratio. To the best of our knowledge, a rigorous analysis of convergence, even in  simple cases, has not been performed for any of the ML algorithms mentioned above. 

The analysis also 
shows that in the specific setting that we consider,   IRAS solves in each iteration a \emph{generalized Rayleigh quotient problem}. This problem plays an important role
in  many  algorithms in data science and ML, including generalized eigenvalue classifiers~\cite{gen_classifiers}, the  common spatial pattern~(CSP) algorithm~\cite{EEG_appl}, and
Fisher's linear discriminant analysis~\cite{book_pattern_class}
(see~\cite{robust_Rayleigh_quotient} for more details). Thus, the  analysis also  creates the first link between
IRAS and  other   ML algorithms. 

  Vectors  (matrices) are denoted by small (capital) letters. For a matrix~$A\in\R^{m\times n}$, $A'$ is the transpose of~$A$.
For a vector~$x\in\R^n$, $x_i \in \mathbb{R}$ is the $i${th} entry of $x$, and
$\|x\|_2:=\sqrt{x'x}$. The vector~$x$ is called   normalized if~$\|x\|_2=1$.
 If~$A\in \R^{n\times n}$ is symmetric, then the ordered eigenvalues of~$A$ are
$
\lambda_1(A)\geq\lambda_2(A)\geq\dots\geq\lambda_n(A),
$
with corresponding eigenvectors $v^1(A),v^2(A),\dots,v^n(A)$.

The remainder of this note is organized as follows. 
Section~\ref{sec:IRAS} reviews the IRAS algorithm. Section~\ref{sec:main} presents the specific setting that we analyze here, and the  main theoretical 
results.  {All the proofs are placed in the Appendix.} 
{Section~\ref{sec:rfmr} describes an application to a model from systems biology.} The final section concludes and describes possible  directions for further research. 

\section{  IRAS algorithm}\label{sec:IRAS}
We briefly review the IRAS algorithm. For more details, see~\cite{IRAS_PNAS}. 
The input to the algorithm is a sequence~$\left\{z^k\right\}_{k=1}^N$. Typically,~$z^k$ is the  {noisy}  $n$-dimensional output 
of a continuous-time dynamical system measured at time~$k\Delta t$. For example, it may be the noisy full state-vector~$z^k=x(k\Delta t)+\eta^k$ , {where~$\eta^k$ is a   noise vector. }

  IRAS 
determines, in an iterative manner, a parametric function~$g:\R^n \times {\R^b} \to\R$. 
The goal is to  
converge to a parameter vector~$\theta^*\in {\R^b}$ such that~$g(z^i;\theta^*)=g(z^j;\theta^*)$ for all times~$i,j$. The parametric function can be linear, that is,~$g(z;\theta)=\theta'z$, polynomial, or, more generally, it may be the output of a neural network, and then~$\theta$ is the set of parameters in this network. 
 Of course, the goal is to find a a parameter vector~$\theta^*$ such that~$g(\cdot;\theta^*) $ is \emph{non-trivial}.

{Since measurements  are always noisy}, 
IRAS  is 
based on   probabilistic
assumptions.  
{The first  assumption 
is that the noisy samples~$z^k$ 
admit a probability density function~(PDF)
 $f_z:\mathbb{R}^{n} \to  \mathbb{R_+}$.}
%
%
For any~$\theta\in {\R^b}$, $z\in\R^n$, and~$x\in\R$,  let 
\begin{equation}
\label{eq:def_fztheta}
f_z^\theta(x):= \int_{\{z|g(z;\theta)=x\}} f_z (z)\diff z. 
\end{equation}
This defines a scalar PDF by projecting with~$g(\cdot;\theta)$. 
Then 
\begin{align}\label{eq:proj}
P(c_1 \leq g(z;\theta) \leq c_2)& = \int_{D^\theta_{c_1,c_2}}f_z(z_1,\dots,z_m)\diff  z_1 \dots  \diff z_m  \nonumber \\
& = \int_{c_1}^{c_2} f_z ^\theta (p)\diff p , 
\end{align}
where $D^\theta_{c_1,c_2}:=\{z \mid c_1 \leq g (z;\theta ) \leq c_2 \}$. In what follows, we   use the notation~$\text{var}(f_z^\theta$) to denote the variance of the {scalar}
random variable~$g(z;\theta).$ From here on, for any PDF $f:\mathbb{R}^n \rightarrow \mathbb{R}_+$ we denote by $f^\theta:\mathbb{R}\rightarrow\mathbb{R}_+$ [$\text{var}(f^\theta)$] the scalar PDF [variance] of its projection 
by~$g(\cdot;\theta)$.

{A natural idea is to find a parameter vector~$\theta^*$ that minimizes~$\text{var}(f^\theta_z)$, but such an approach usually leads to a trivial solution. For example, if~$g(z;\theta)=\theta'z$ this may lead to the trivial solution~$\theta^*=0$.  }

To find  a non-trivial~$\theta^*$, 
IRAS  uses also 
a surrogate PDF~$f_{\bar z }:\R^n \to \R_+$,
{corresponding to a random variable~$\bar z$}. 
Roughly speaking, the idea is then to find 
a vector~$\theta$ that minimizes the ratio
\begin{equation}\label{eq:ratiof}
\frac{ \mathrm{var} ( f_z^{\theta } ) }{ \mathrm{var} ( f_{\bar z}^{\theta } ) } =
\frac{\mathrm{var} (  g(z^k;\theta) ) }
{\mathrm{var} ( g(\bar z^k;\theta) )} .
\end{equation}
{Then IRAS will not converge to a  trivial~$\theta$ as this will give  a ratio of~$0/0$ in~\eqref{eq:ratiof}. }

The algorithm initializes~$f_{\bar z}$
such that  the following 
weak requirement holds
\begin{align}
 \label{eq:supp}   
&\text{supp}(f_z )\subseteq \text{supp} (f_{\bar z}), \text{ and } \nonumber \\  &0<\int_D f_z(z)\diff z \not = \int_D f_{\bar  z} (z)\diff z
\text{ for at least one set } D  \subseteq \R^n.
\end{align}


\begin{Remark}\label{rem:updat_surr}
Instead of using a constant PDF~$f_{\bar z}$,
 IRAS updates the surrogate PDF at each iteration. To explain why, assume that there exists a set~$D\subseteq\R^n$ such that 
\begin{align*}
  \int_D f_{\bar{z}}(q) \diff q > 0, \text{ and } \int_D f_{z}(q) \diff q = 0 
\end{align*}
then  any vector $\theta$ for which~$g(z;\theta) = 0$ for all $z \notin D$ and~$g(z;\theta) \neq 0$ for $z \in D$, 
minimizes the ratio in~\eqref{eq:ratiof}  (achieving a value  zero). Updating the surrogate PDF is needed to avoid such trivial solutions for~$\theta$. 
\end{Remark} 

  IRAS  iterates between two steps. At iteration~$i$ the algorithm already has an estimate~$\theta^{i-1}$ for the parameter vector in~$g$, and the two steps are:  \\
 {\bf Step 1.} Update the current surrogate PDF~$f_{\bar z}$ to a surrogate PDF {$f_{\tilde z}:\R^n \times\R^b \to\R_+$} by
\begin{equation}\label{eq:advers}
f_{\tilde z}(q;\theta^{i-1}):=f_{\bar z}(q) \zeta\left(g(q;\theta^{i-1} );\theta^{i-1}\right),
\end{equation}
where~$\zeta:\R    \times\R^b \to\R_+$ is a weighting function that depends on the current estimate~$\theta^{i-1}$. 
The weighting function is chosen such that {the modified surrogate PDF satisfies}
\begin{equation}\label{eq:fc_equal_f}
      f^{\theta^{i-1}}_{\tilde z }(p;\theta^{i-1})=   f_z^{\theta^{i-1}}(p)  \text{ for almost all } p \in\R  . 
\end{equation}
 { In other words, the goal is to shift~$f_{\bar z}$ 
to  a PDF~$f_{\tilde z}$ such that the projection of~$f_{\tilde z}$  and~$f_z$ 
under~$g(\cdot;\theta^{i-1})$ are identical. }This  overcomes  the difficulty described in Remark~\ref{rem:updat_surr}.
A closed-form expression for~$\zeta$ 
 is given in Lemma~\ref{lem:zeta} below.

{\bf Step 2.} 
Update  the estimate of the vector of parameters by setting 
\begin{equation} \label{eq:ration}
\theta ^{i}:=  \argmin_\theta  \frac{ \text{var} (f_z^\theta)   }{\text {var} (f_{\tilde z}^\theta  (\theta^{i-1}  ) )   }.
\end{equation} 

Note that~\eqref{eq:fc_equal_f}
implies that
\[
\left .
  \frac{ \text{var} (f_z^\theta  ) )   }{\text {var} (f_{\tilde z}^\theta  (\theta^{i-1}  ) )   } \right  |_{\theta=\theta^{i-1}} =1 , 
\]
 so the first step of the algorithm given in \eqref{eq:advers} may be regarded as an ``adversarial''
 step, in the sense that it is ``opposite'' to the second step in~\eqref{eq:ration} that minimizes the ratio~$\text{var} (f_z^\theta)/\text{var}(f^\theta_{\tilde z})$.

The iterations continue until the difference between~$\theta^{i-1}$ and~$\theta^i$ is sufficiently small. 
The IRAS algorithm has been introduced in the context of identifying
what is actually  regulated (i.e., what is the invariant function~$g$) in biological systems, 
hence the terminology 
{\sl Identifying Regulation with Adversarial
Surrogates Algorithm}. 

   \section{Main Results}\label{sec:main}
   In this section, we rigorously analyze   the behaviour of~IRAS in a  specific setting. 
   
For the analysis, we require the following fact.
\begin{Lemma} \label{lem:zeta}
\cite{IRAS_PNAS}
The function~$\zeta $ 
guaranteeing  that~\eqref{eq:fc_equal_f}
 holds is given by
 \begin{equation}
     \zeta(x;\theta)=\begin{cases} 
     f_z^\theta (x ) /f_{\bar z}^\theta   (x  ),& \text{ if } f_{\bar z}^\theta   (x  ) \not =0,\\
     0,& \text{otherwise}.
     \end{cases}
 \end{equation}
Furthermore, this~$\zeta$ also guarantees that~$f_{\tilde z}(q;\theta^{i-1})$ in
\eqref{eq:advers} is indeed a PDF. 
 \end{Lemma}
 Note that this provides a closed-form expression for the weighting  function~$\zeta$ used in Step~1 of~IRAS. 

The next result addresses
the case of a trivial solution {described}
in Remark~\ref{rem:updat_surr}.
\begin{Proposition}\label{prop:prop1}
    Suppose that: (a)~there exists a set~$D$  that is not a subset of~$ \mathrm{supp}(f_z)$ for which 
    \begin{align*}
        f_{\tilde{z}}(q;\theta^{i-1}) > 0 \text{ for all } q \in D, \text{ and } \int_{q \in D} f_z(q) \diff q = 0;
    \end{align*}
    and (b)~$\theta^i$ is a trivial minimizer
of the optimization problem in Step~$2$, i.e. it  satisfies~$g(z;\theta^i) = 0$ for all~$z \notin D$ and~$g(z;\theta^i) \neq 0$ for all $z \in D$. Then
    \begin{align*}
        \int_{q \in D} f_{\tilde{z}}(q;\theta^{i}) \diff q=0.
    \end{align*}
   \end{Proposition}
Note  that this implies
 that the $i$th iteration
 of IRAS  eliminated the trivial solution~$\theta^{i-1}$.

%
%
%

We now analyze    IRAS 
in the  case where the  {$n$-dimensional}   output   
admits a linear first integral ({so the output belong to a subspace}), 
and the  measurements 
include Gaussian noise.  

\subsection{The model}
Fix a normalized vector~$v^1\in\R^n$.
Consider  the  subspace~$H \subset \R^n$ defined by 
$
  H :=\{x \in \mathbb{R}^n \mid (v^{1})' x=0\} . 
$
Fix normalized vectors~$v^2,\dots,v^n$ such that~$\{v^1,\dots,v^n\}$ is an orthonormal basis of~$\R^n$. Then 
$
H =  \left\{\sum_{i=2}^n c_i v^i \mid c_i \in \mathbb{R}\right\} . 
$

Any vector~$z \in \mathbb{R}^n$ can be decomposed  as 
\begin{equation}
    \label{eq:dec_z}
z= \sum_{i=1}^n c_i(z) v^i,\text{ with }c_i(z):=z' v^i.
\end{equation}
Thus,~$z\in H$ if and only if (iff)~$c_1(z)=0$.

Let~$\mathcal{N}_n (x|\mu,G) $ denote the $n$-dimensional
Gaussian distribution
with mean~$\mu\in\R^n$ and
 positive-definite covariance
 matrix~$G \in\R^{n\times n} $.
 We assume that the invariant   is that the measurements are   in~$H$ {i.e., they satisfy a linear first integral}, but are contaminated with noise. 
Specifically,  the measurements~$z\in\R^n$, decomposed    in the form~\eqref{eq:dec_z}, are drawn  
 according to the~PDF 
 \begin{equation}\label{eq:f_z_z}
 f_z(z) =  \mathcal{N}_1 ( c_1(z) |0,\sigma^2)  \prod_{i=2}^n \mathcal{N}_1 (  c_i(z)  | 0,1 ),
 \end{equation}
 with~$\sigma\in(0,1)$. 
The term~$\prod_{i=2}^n \mathcal{N}_1 (  c_i(z)  | 0,1 )  $ 
corresponds to points in~$H$. The Gaussian distribution with zero average   favors smaller  values of~$c_i(z)$, corresponding to the fact that typically the outputs tend to be bounded. The term~$  \mathcal{N}_1 ( c_1(z) |0,\sigma^2) $ corresponds to  noise in the direction  that is  perpendicular to~$H$. The assumption  that~$\sigma\in(0,1)$ implies that  the variance of the noise is smaller than the variance of the signal. 

We assume that~IRAS
 searches over the family of linear functions
 \begin{equation}\label{eq:g_z_theta}
    g(z; \theta) = \theta' z, 
\end{equation}
with parameter vector~$\theta \in \mathbb{R}^{n}$,
and that it initializes  the surrogate PDF by
 \begin{equation*}
    f_{\bar{z}}(\bar z) = \mathcal{N}_n(\bar{z}; 0, \bar \Sigma ), \text{ with } \bar \Sigma=\bar \sigma^2 I_n,  \; \bar\sigma>0.
\end{equation*}
Note  that this implies that~\eqref{eq:supp} holds. This completes the description of the IRAS setting.
We now analyze    IRAS for this case. 
\subsection{Analysis }
\begin{Proposition}\label{prop:ratio}
Define a matrix~$\Sigma\in\R^{n\times n}$ as the inverse of the matrix
\begin{align}\label{eq:def_sigma_minus_one}
\Sigma^{-1}&:= \sigma^{-2} v^1 (v^1)' +v^2 (v^2)'+\dots+v^n (v^n)'   .
 \end{align}
For~$\theta\in\R^n$, let 
    \[
s(\theta):=\theta' \Sigma \theta,   \quad \bar s(\theta):=\theta' \bar \Sigma \theta ,
\]
and
\begin{align}\label{eq:Sherman}
\tilde \Sigma (\theta )  
                &:= \bar \Sigma  -\left ( (\bar s(\theta))^{-1} - (\bar s(\theta))^{-2} s(\theta) \right ) \bar \Sigma \theta \theta' \bar \Sigma.
\end{align}
Under the assumptions specified above, the $i$th iteration 
of~IRAS yields
\begin{equation}\label{eq:gauss_ratio} 
\theta ^i   
= \argmin_\theta  \frac{ \theta' \Sigma \theta   }{\theta' \tilde{\Sigma}(\theta^{i-1}) \theta}. 
\end{equation} 
%
\end{Proposition}
Note that for any scalar~$c\in \R\setminus\{0\}$ we have
$
 \frac{ (c\theta)' \Sigma (c\theta)   }{(c\theta)' \tilde{\Sigma}(\theta^{i-1}) (c\theta)} = 
 \frac{ \theta' \Sigma \theta   }{\theta' \tilde{\Sigma}(\theta^{i-1})\theta}  , 
$
so we may assume, without loss of generality (WLOG), that~$\|\theta^i\|_2=1$ for all~$i$. This implies in particular  that~IRAS  does not converge to the trivial solution corresponding to~$\theta=0$.

 The right-hand side in~\eqref{eq:gauss_ratio} 
is a generalized Rayleigh quotient, and the iteration   generating~$\theta^i$ from~$\theta^{i-1}$
is widely used. In
computational physics and chemistry it is known as the self-consistent-field~(SCF)
iteration (see, e.g.,~\cite{comp_chem}).
However, as noted in~\cite{robust_Rayleigh_quotient}, its convergence behavior  is  
not well-understood.

 It is 
 well-known~\cite{ghojogh2019eigenvalue,robust_Rayleigh_quotient} that the solution~$\theta^i$ of~\eqref{eq:gauss_ratio}
is the eigenvector corresponding to  the minimal  eigenvalue~$\lambda$ in the 
 generalized eigenvalue problem:
\[
\Sigma \theta^i =\lambda \tilde\Sigma(\theta^{i-1 }) \theta^i . 
\]
 Equivalently,
 \begin{equation}\label{eq:domi}
 \lambda^{-1} \theta^i=  \Sigma^{-1} \tilde \Sigma(\theta^{i-1 }) \theta^i ,
 \end{equation}
 so~$\theta^i$ is the eigenvector of~$\Sigma^{-1} \tilde \Sigma(\theta^{i-1 }) $ corresponding to the largest eigenvalue of~$\Sigma^{-1} \tilde \Sigma(\theta^{i-1 }) $ (i.e., the dominant eigenvector of~$\Sigma^{-1} \tilde \Sigma(\theta^{i-1 }) $).

Eq.~\eqref{eq:domi} implies that~$\theta^e\in\R^n$ is an
equilibrium    of IRAS
iff  it   
is the dominant
eigenvector 
 of~$ \Sigma^{-1} \tilde \Sigma(\theta^{e }) $.
To determine when this happens, we first
analyze the spectral properties of~$\Sigma^{-1} \tilde \Sigma(\theta)$.
\begin{Lemma}\label{lemma:theta_perp}
Fix a normalized vector~$\theta\in\R^n$.
Let~$\theta_\perp\in\R^n$ be a normalized  vector such that~$\theta'\theta_\perp=0$. Then
\begin{align}\label{eq:effct_theta_perp}
    \Sigma^{-1}\tilde\Sigma(\theta)\theta&=s(\theta) \Sigma^{-1}\theta,\nonumber\\
    \Sigma^{-1}\tilde\Sigma(\theta)\theta_\perp&=\bar\sigma^2 \Sigma^{-1}\theta_\perp.
\end{align}
\end{Lemma}
%
%
\begin{Remark}\label{remark:perp}
    The proof of Lemma~\ref{lemma:theta_perp} {(see the Appendix)} implies that~$\det(\tilde\Sigma(\theta))=s(\theta)\bar\sigma^{2(n-1)} $. Note also  that if~$\theta_\perp=\sum_{i=2}^ n d_i v^i$, with~$d_i\in\R$, then~\eqref{eq:effct_theta_perp} gives~$\Sigma^{-1}\tilde \Sigma(\theta)\theta_\perp=\bar\sigma^2 \theta_\perp$, i.e.
    $\theta_\perp$ is an eigenvector of~$\Sigma^{-1}\tilde \Sigma(\theta)$ corresponding to the eigenvalue~$\bar\sigma^2$.
\end{Remark}
The next result  analyzes the steady-state of~IRAS, and provides a condition guaranteeing  that the     vector~$v^1$ is indeed an  equilibrium of the iterative algorithm.
\begin{Theorem}\label{thm:main_bar_sigma}
Under the assumptions specified above,
the vector~$v^1$ is a steady-state of~IRAS iff~$\bar \sigma<1$.
\end{Theorem}
%
%
\begin{Remark}\label{rem:H_TRANSP}
    Let~$V:=\begin{bmatrix}
    v^1&\dots&v^n
    \end{bmatrix}$.  Note that~$V'=V^{-1}$. By 
   the proof of Thm.~\ref{thm:main_bar_sigma} {(see the Appendix)},
$
V'  \Sigma^{-1}\tilde\Sigma(v^1) V =\diag(1,\bar \sigma^2,\dots,\bar \sigma^2).  
$
Taking transpose  yields 
$
V'  (  \Sigma^{-1}\tilde\Sigma(v^1) )' V =\diag(1,\bar \sigma^2,\dots,\bar \sigma^2) ,  
$
implying in particular that~$\Sigma^{-1}\tilde\Sigma(v^1)$ is symmetric. 
\end{Remark}

The next step is to analyze convergence of IRAS to the   vector~$v^1$, {that is, to the linear first integral}. The next result 
shows that for some initial conditions   IRAS  
converges to~$v^1$ in a single iteration.

\begin{Proposition}\label{prop:firstIter}
    Suppose that
    \begin{equation}\label{eq:sig_sig_cond}
    \sigma^2 <\bar\sigma^2<1.
    \end{equation}
    If the  initial condition is~$\theta^0=
\sum_{i=2}^n d_i v^i$, with~$d_i\in\R$, then the first iteration of  IRAS yields~$\theta^1=v^1$.
\end{Proposition}
%
%
The next result shows that, at least in some directions, IRAS 
locally converges  to the desired value~$v^1$. 
\begin{Proposition}\label{prop:2cond}
Suppose that  condition~\eqref{eq:sig_sig_cond} holds and, in addition, 
\begin{align}\label{eQ:add_mo}
    2 \bar\sigma^2< 1+\sigma^2 .
\end{align}
Assume that~$\theta^0$ is proportional  to~$ v^1+\varepsilon v^k  $, with~$k\in\{2,\dots,n\}$,
and~$\varepsilon\in\R$. There exists~$\eta>0$ such that for any~$|\varepsilon|<\eta$  IRAS converges to~$v^1$. 
\end{Proposition}
%

{\section{An Application}\label{sec:rfmr}
The ribosome flow model on a ring~(RFMR)  
has been  used to analyze  the flow of ribosomes along   a circular mRNA molecule during  translation~\cite{RFMR,RFMR_OPTIMAL}. 
In this nonlinear dynamical model, the mRNA molecule is divided into~$n$ sites, where each site corresponds to   a set of consecutive codons along the mRNA. The state-variable~$x_i(t)$,  $i=1,\dots,n$, describes the density of ribosomes  at site~$i$ at time~$t$. The densities are normalized so that~$x_i(t)\in[0,1]$ for all~$t$, with~$x_i(t)=0$ [$x_i(t)=1$] implying that the site is empty [completely full]. 
The flow of ribosomes from site~$i$ to site~$i+1$ is~$\lambda_ix_i(1-x_{i+1})$, i.e. the flow  increases with the density at site~$i$, and decreases with the density at the next site~$x_{i+1}$. 
The parameter~$\lambda_i>0$ is the transition rate from site~$i$ to site~$i+1$ ($\lambda_i$ 
is proportional to the reciprocal of
the decoding time of the ribosome at site~$i$). 
The RFMR equations are thus
\begin{align*}
\dot{x}_i &= \lambda_{i-1} x_{i-1} (1-x_i) - \lambda_{i} x_{i} (1-x_{i+1}),\quad i=1,\dots,n,
\end{align*}
with all indexes modulo~$n$ (for example,~$x_0=x_n$ and~$x_{n+1}=x_1$). Thus,  the change in the density at site~$i$  is the     flow from site~$i-1$ to site~$i$ minus the flow  from site~$i$ to site~$i+1$.  

It follows from these equations that~$\sum_{i=1}^n\dot x_i(t)\equiv 0$, so~$H(x):=1_n' x$ is a linear first integral of the dynamics, where~$1_n\in\R^n$ is a vector with all entries one. This represents the fact that ribosomes do not enter or leave the circular chain, so their total density is conserved. 

We applied   IRAS   to this model\footnote{The code for the simulations is  available at \href{https://github.com/RonTeichner/IRAS/blob/main/IRAS_Ribosome_RQ_paper.ipynb}{https://github.com/RonTeichner/IRAS}} with (the arbitrarily chosen) parameters~$n=5$, 
$\lambda=\begin{bmatrix}
    2&5&5&0&1
\end{bmatrix}'$ and
  initial condition $x(0) = 
\begin{bmatrix}
    0.71 &0.9&  0.28& 0.8&  0.76
\end{bmatrix}'$.
The data used by IRAS consists of noisy
samples of the  state vector~$z^k = x(0.001*k) + \eta^k$, $k=1,\dots,N$, with~$N=2000$,
and~$\eta^k \sim \mathcal{N}_5(0,\sigma^2 I_5)$, $\sigma^2=10^{-5}$. The surrogate distribution was initialized to~$f_{\bar z} = \mathcal{N}_5(\mu, \bar \Sigma)$ where $\mu: = \frac{1}{N}\sum_{k=1}^{N} z^k$, and $\bar \Sigma = \bar \sigma^2 I_5$. 

  Props.~\ref{prop:firstIter} and~\ref{prop:2cond} suggest  that setting $\bar \sigma^2  = 2 \times 10^{-5}$, i.e. slightly higher than~$\sigma^2$, would lead to convergence. Iterating IRAS for this value of~$\bar \sigma^2$, with the initial condition~$\theta^0=\begin{bmatrix} -0.12 & 0.20 & 0.41 & 0.76 & 0.45\end{bmatrix}'$ gives the sequence of normalized
vectors: 
\begin{align*}
    \theta^1&=\begin{bmatrix} 0.43 & 0.45 & 0.43 & 0.47& 0.45 \end{bmatrix}' , \\
    \theta^2&=\begin{bmatrix} 0.46 & 0.44 & 0.44 & 0.45 & 0.44 \end{bmatrix}' , 
%
\end{align*}
where all numerical values are to 2 digit accuracy. Thus, the algorithm  quickly  converges to a solution proportional to~$1_5$.
%
%
 }
 \section{Discussion}
ML algorithms have achieved considerable empirical success, providing efficient solutions to complex problems across various domains. However, their theoretical understanding is still lacking. Here, we analyzed    
the recently suggested  IRAS algorithm in a specific setting of measurements that admit a linear invariant up to Gaussian noise. We showed that in this case IRAS   minimizes   a generalized Rayleigh  quotient at each iteration, linking it  to several other important~ML algorithms. We proved  that the  linear  invariant is an
equilibrium point of IRAS, and that it is locally stable, at least with respect to some perturbation directions. 

{Topics for further research include studying   IRAS  when the data satisfies a  non-linear  invariant, e.g.   those that are a linear combination of known, nonlinear, basis functions.
Another potential research  direction is to use
the theoretical analysis to improve
the algorithm. }

{\subsubsection*{Acknowledgements} We thank the editor and the anonymous referees for their detailed and helpful comments. }

\section*{Appendix: Proofs}
\emph{Proof of Lemma~\ref{lem:zeta}}.
Note that
\begin{align*}
        f^\theta_{\tilde{z} }( x ; \theta) &=  \int_{\{z \mid g(z;\theta)=x\}} f_{\tilde{z}}(z;\theta) \diff z\\ 
        &=  \int_{\{z \mid g(z;\theta)=x\}} f_{\bar{z}}(z)\zeta(g(z;\theta)) \diff z\\ 
        &=  \int_{\{z \mid g(z;\theta)=x\}} f_{\bar{z}}(z)\zeta(x)\diff z \\ 
        &=  \zeta(x) \int_{\{z \mid g(z;\theta)=x\}} f_{\bar{z}}(z)\diff z \\ 
        &=  \zeta(x) f_{\bar{z}}^{\theta}(x)\\ 
        &=  (  f_z^\theta (x   )/ f_{\bar z}^\theta   (x  ))  f_{\bar{z}}^{\theta}(x)=f_z^\theta (x ).
        %
\end{align*}
 Note that~\eqref{eq:supp} 
 implies that the  ratio~$  f_z^\theta (x )/f_{\bar z}^\theta   (x  ) $ is well-defined. 
We now show   that~$f_{\tilde z}(q;\theta^{i-1})$ in
\eqref{eq:advers} is indeed a~PDF. Note that 
\begin{equation}\label{eq:tile_z_comp}
    \begin{split}
        \int f_{\tilde{z}}(q;\theta  ) \diff q
        &= \int f_{\bar{z}}(q  )\zeta(g(q; \theta))\diff q\\&
        = \int_{-\infty}^\infty \diff x \int_{\{q \mid g(q;\theta)=x\}} f_{\bar{z}}(q)\zeta(g(q;\theta))\diff q\\
        &= \int_{-\infty}^\infty  \zeta(x) \diff x \int_{\{q\mid g(q;\theta)=x\}} f_{\bar{z}}(q)\diff q\\&
        =\int_{-\infty}^\infty \frac{f_z^\theta (x ) }{f_{\bar z}^\theta   (x  )} \diff x f_{\bar z}^\theta   (x  )\\
        &=\int_{-\infty}^\infty  f_z^\theta (x ) \diff x=1,
    \end{split}
\end{equation}
and this completes the proof.~\hfill{\qed}

\emph{{Proof of Prop.~}\ref{prop:prop1}}.
    Note that $g(z;\theta^i)=0$ for all $z \in \text{supp}(f_z)$, yielding $f_z^{\theta^i}(p)=\delta(p)$, where $\delta$ is the Dirac delta function. 
    By Lemma~\ref{lem:zeta},
    Step~1 of the next iteration yields 
\begin{align*}
    \zeta\left(g(q;\theta^{i} );\theta^{i}\right)  &= \frac{f_z^{\theta^i} (g(q;\theta^{i} ) ) }{f_{\bar z}^{\theta^i}   (g(q;\theta^{i} )  )}
    = \frac{\delta (g(q;\theta^{i} ) ) }{f_{\bar z}^{\theta^i}   (g(q;\theta^{i} )  )} .
\end{align*}  
This implies that for any~$q \in D$, we have 
$\zeta\left(g(q;\theta^{i} );\theta^{i}\right) = 0$,
as  $\theta^i$ satisfies $g(q;\theta^i) \neq 0$ for all $q \in D$. It follows that the updated surrogate PDF satisfies 
\begin{align*}
    \int_{q \in D} f_{\tilde{z}}(q;\theta^{i}) \diff q
    &= \int_{q \in D} f_{\bar{z}}(q)\zeta\left(g(q;\theta^{i} );\theta^{i}\right) \diff q =0,
\end{align*}
and this completes the proof.~\hfill{\qed}
%
%

\emph{{Proof of Prop.~}\ref{prop:ratio}}.
%
Note that
\begin{align}\label{eq:fzz_fir}  
f_z(z)& = 
c_z
\exp( -\frac{1}{2} (  \sigma^{-2}c_1 ^2(z)+c_2^2(z)+\dots+c_n^2(z)  )  )\nonumber\\
&=  
c_z
\exp( -\frac{1}{2}z' \Sigma^{-1}  z),
\end{align}
with $c_z:=\frac{1}{ (\sqrt{2\pi})^{n}\sqrt{ \sigma^2}}$, and~$\Sigma^{-1}$ in~\eqref{eq:def_sigma_minus_one}.
 The eigenvalues of~$\Sigma^{-1}$ 
 are~$\lambda_1=\sigma^{-2}$, $\lambda_2=1,
 \dots,\lambda_n=1$, with corresponding eigenvectors~$v^1,v^2,\dots,v^n$,
so~$\det(\Sigma^{-1})=\sigma^{-2}$. 
Hence,~$\det(\Sigma)=\sigma^2$ , and 
we can write~\eqref{eq:fzz_fir} as
$
f_z(z)= \frac{1}{ (\sqrt{2\pi})^{n} \sqrt{\det(\Sigma )} }
\exp(-\frac{1}{2} z' \Sigma ^{-1} z  ),
$
i.e., a  Gaussian distribution with average zero and covariance matrix~$\Sigma$. Note that
$
\Sigma =  \sigma^2v^1(v^1)'+v^2(v^2)'+\dots+ v^n (v^n)'.
$

  By~\eqref{eq:g_z_theta},  $\text{var}(f_z^\theta)=\theta' \Sigma  \theta$,
%
%
and  Lemma \ref{lem:zeta}
gives
\begin{align*} 
    \zeta(x;\theta) &= \frac{f_z^\theta (x ) }{f_{\bar z}^\theta   (x  )} 
    = \frac{\frac{1}{\sqrt{2\pi\theta' \Sigma  \theta}}
    \exp (-\frac{1}{2}x'(\theta'\Sigma  \theta)^{-1}x) }{\frac{1}{\sqrt{2\pi\theta' \bar \Sigma  \theta}}
    \exp(-\frac{1}{2}x'(\theta' \bar \Sigma  \theta)^{-1}x) } \\
    &= \sqrt{\frac{\bar s( \theta) }{s( \theta)}}
    \exp( -\frac{1}{2}x'\left( (s( \theta))^{-1} - (  \bar 
   s( \theta))^{-1}\right)x ) .
\end{align*}
%
Using~\eqref{eq:advers} gives 
\begin{align*}
f_{\tilde z}(q;\theta^{i-1})&= f_{\bar z}(q) \zeta( q' \theta^{i-1} ;\theta^{i-1}) \\
&=c \exp(-\frac{1}{2} q'  (\tilde \Sigma (\theta^{i-1})  )^{-1} q )  ,
\end{align*}
with~$c:= \frac{1}{2 \pi  \sqrt{\det(\bar \Sigma )  }}\sqrt{\frac{\bar s( \theta) }{s( \theta)}}$,
 and
 \[
\tilde \Sigma (\theta) =\bigl( (\bar\Sigma)^{-1}+ ((s( \theta))^{-1} - (  \bar 
   s( \theta))^{-1}) \theta (\theta)'\bigl)^{-1}. 
\]

%
Lemma \ref{lem:zeta}   guarantees that $f_{\tilde{z}}(\theta)$ is a PDF, implying that it is Gaussian with zero average and covariance matrix~$\tilde \Sigma(\theta^{i-1})$. 
The Sherman-Morrison-Woodbury formula~\cite[Ch.~0]{matrx_ana} gives~\eqref{eq:Sherman}. %
%
 %
Since in the 
 $i${th} iteration,  IRAS sets
$  \theta ^i  =  \argmin_\theta  \frac{ \text{var} (f_z^\theta )   }{\text {var} (f_{\tilde z}^\theta  (\theta^{i-1}  ) )   },
$
 this completes the proof.~\hfill{\qed}

\emph{{Proof of Lemma~}\ref{lemma:theta_perp}}.
%
Since~$\bar\Sigma=\bar\sigma^2 I_n$, we have~$\bar s(\theta)=\theta'\bar \Sigma\theta =\bar\sigma^2 \theta '\theta=\bar\sigma^2$,    
and~\eqref{eq:Sherman} gives
\begin{align}\label{eq:tilde_sigma}
\tilde \Sigma(\theta) &= 
\bar\sigma^2 I_n -\left (  \bar\sigma^{-2}  -  (\theta)'\Sigma\theta \bar\sigma^{-4} \right ) \bar\sigma^4 \theta  \theta ' \nonumber \\
&= \bar\sigma^2 I_n  -\left ( \bar\sigma^2  -  \theta'\Sigma\theta
\right )   \theta  \theta '
.
\end{align}
Hence,
$
\tilde \Sigma(\theta)\theta  = 
   s(\theta) \theta   ,
$
and for any vector~$\theta_\perp$ that is orthogonal to~$\theta$, we have 
$
\tilde \Sigma(\theta)\theta _\perp= \bar\sigma^2
\theta_\perp
$.~\hfill{\qed}

\emph{{Proof of Thm.~}\ref{thm:main_bar_sigma}}.
%
By Lemma~\ref{lemma:theta_perp},
$
    \Sigma^{-1} \tilde\Sigma(v^1) v^1 
     = v^1$,
and for any~$i>1$, we have
$    \Sigma^{-1} \tilde\Sigma(v^1) v^i  = \bar\sigma^2 v^i$.
We conclude that~$1$
 is a simple dominant eigenvalue of~$ \Sigma^{-1} \tilde\Sigma(v^1)$ iff~$\bar \sigma<1$. The corresponding eigenvector is~$v^1$.~\hfill{\qed}

\emph{{Proof of Prop.~}\ref{prop:firstIter}}.
 Let~$d:=\begin{bmatrix}
     d_2&\dots& d_n
 \end{bmatrix}' \in\R^{n-1}$.
Since we may assume that~$\theta_0$ is normalized,~$d'd=1$. 
Fix $n-2$ linearly independent and normalized  vectors~$p^1,\dots,p^{n-2} \in \R^{n-1}$
such that~$(p^k)'d =0$ for all~$k\in\{1,\dots,n-2\}$. Then the vector~$w^k:=\sum_{i=2}^n p^k_{i-1} v^i \in\R^n$ is orthogonal to~$\theta^0$ for all~$k\in\{1,\dots,n-2\}$.
By Lemma~\ref{lemma:theta_perp},     \begin{align*}
       \Sigma^{-1} \tilde\Sigma(\theta^0)\theta^0 = \theta^0, \quad
\Sigma^{-1} \tilde\Sigma(\theta^0) v^1
         =\sigma^{-2}\bar \sigma^2  v^1,
       \end{align*}
 and 
\[
\Sigma^{-1} \tilde\Sigma(\theta^0) w^k        = \bar \sigma^2  w^k, \quad k \in \{ 1,\dots,n-2   \} .
\]
Thus, the eigenvalue-eigenvector pairs of~$  \Sigma^{-1} \tilde\Sigma(\theta^0) $ are: 
$
(\sigma^{-2}\bar\sigma^2 ,v^1), (1,\theta^0), (\bar\sigma^2 ,w^1),\dots,(\bar\sigma^2 , w^{n-2}). 
$
Eq.~\eqref{eq:sig_sig_cond} implies that~$v^1$ is the dominant eigenvector of~$  \Sigma^{-1} \tilde\Sigma(\theta^0)$.~\hfill{\qed} 

\emph{{Proof of Prop.~}\ref{prop:2cond}}.
Assume, WLOG, that~$k=2$, and that~$\theta^0$ is normalized, 
that is,~$\theta^0=\frac{v^1+\varepsilon v^2}{(1+\varepsilon^2)^{1/2}}$.
The  proof is based on determining the eigenvalues and eigenvectors of~$\Sigma^{-1} \tilde \Sigma(\theta^0)$. 
By Remark~\ref{remark:perp},
$(\bar\sigma^2,v^k)$ is an eigenvalue-eigenvector pair of~$ \Sigma^{-1} \tilde \Sigma(\theta^0)$ for any~$k\in\{3,\dots,n\}$. 
Our goal is to determine the other two eigenvalues denoted~$\lambda_1,\lambda_2$.

%
Note that
\begin{align}\label{eq:effect_theta_zeo}
    \Sigma^{-1} \tilde \Sigma(\theta^0) \theta^0
    &= (\sigma^2+\varepsilon^2) (1+\varepsilon^2)^{-3/2} (\sigma^{-2}v^1+\varepsilon v^2). 
\end{align}
The normalized 
vector~$\theta^0_\perp:=\frac{\varepsilon v^1- v^2}{(1+\varepsilon^{2})^{1/2}}$ is orthogonal to~$\theta^0$, and using~\eqref{eq:effct_theta_perp} gives  
\begin{align}\label{eq:effect_theta_zer0_prep}
    %
    %
    %
    \Sigma^{-1} \tilde \Sigma(\theta^0) \theta^0_\perp & =  \bar\sigma^2  (1+\varepsilon^{2})^{-1/2}  (\varepsilon\sigma^{-2} v^1- v^2). 
\end{align}
Our goal now is to find an eigenvector  of~$ \Sigma^{-1} \tilde \Sigma(\theta^0) $  in the form~$\theta^0+ c \theta^0_\perp$, that is, to determine~$\lambda,c\in\R$ such that
\[
 \Sigma^{-1} \tilde \Sigma(\theta^0) (\theta^0+ c \theta^0_\perp) =\lambda  (\theta^0+ c \theta^0_\perp).
\]
Using~\eqref{eq:effect_theta_zeo} and~\eqref{eq:effect_theta_zer0_prep}, 
and comparing the coefficients of~$v^1$ and~$v^2$ yields the two equations:
\begin{align*}
                 (\sigma^2+\varepsilon^2) \sigma^{-2}
                 (1+\varepsilon^2)^{-1}    +
                 c \varepsilon \bar \sigma^2 \sigma^{-2} 
                &= 
  (1+\varepsilon c )\lambda   ,\\
                  (\sigma^2+\varepsilon^2) \varepsilon  (1+\varepsilon^2)^{-1}    -
               c \bar \sigma^2     &= 
 \lambda (\varepsilon-c)   .
\end{align*}
%
%
%
The first equation yields 
\begin{equation}\label{eq:def_c_ev}
c(\lambda)=
\frac{ (\sigma^2+\varepsilon^2)\sigma^{-2}-\lambda(1+\varepsilon^2)  }
{ \varepsilon(1+\varepsilon^2) (\lambda-\bar\sigma^2\sigma^{-2})},
\end{equation}
and substituting this in the second equation and simplifying 
gives
$
a_1 \lambda^2 +a_2\lambda+
a_3=0 $,
with
 $
    a_1  :=( 1+  \varepsilon^2)^2$,
  $a_2  := -  (\sigma^2+\varepsilon^2) ( \varepsilon^2+\sigma^{-2} +(1+\varepsilon^2) \bar\sigma^2 \sigma^{-2} ) $,  and 
    $a_3 :=( \sigma^{2}+\varepsilon^2 )(1+\varepsilon^2)\sigma^{-2} \bar \sigma^{2}.
$
%
Solving this quadratic equation gives the two eigenvalues 
\begin{align}\label{eq:lam1_lam2}
\lambda_{1}=
    \frac{-a_2 + \sqrt{\Delta}}
    {2   \left(1+\varepsilon ^2 \right)^2},\quad
     \lambda_{2}=
    \frac{-a_2 - \sqrt{\Delta}}
    {2   \left(1 +\varepsilon ^2\right)^2},
\end{align}
with
%
%
%
$\Delta :=  a_2^2-4a_1a_3$. 

For~$\varepsilon=0$, we get $a_2 \mid_{\varepsilon=0} =  -  1  - \bar\sigma^2$ and $\Delta  \mid_{\varepsilon=0} =
      (1-\bar\sigma^2 )^2$,
so~\eqref{eq:sig_sig_cond} implies
that~$\Delta  >0$ for any~$|\varepsilon|$ sufficiently  small. Also,~$\lambda_1\mid_{\varepsilon=0}=1$. Thus,~$\lambda_1,\lambda_2$ are real,  and since~$\lambda_3=\dots=\lambda_n=\bar\sigma^2$, we conclude that~$\lambda_1$ is the dominant  and simple eigenvalue
for any~$|\varepsilon|$ sufficiently  small.
Using~\eqref{eq:def_c_ev}
implies that the normalized dominant eigenvector is
\begin{align}\label{eq:w_def}
    w &:=\frac{\theta^0+c(\lambda_1)\theta^0_\perp}{\|\theta^0+c(\lambda_1)\theta^0_\perp\|_2} \nonumber \\
    &= \frac{  (1+\varepsilon c(\lambda_1))v^1+(\varepsilon-c(\lambda_1) )v^2  }{  \sqrt {   (1+\varepsilon c(\lambda_1))^2+ (\varepsilon-c(\lambda_1) )^2    } } .
\end{align}
Expanding this as a Taylor series in~$\varepsilon$ yields
$ w 
= v^1 +  r \varepsilon v^2 + o(\varepsilon)$,  with
$
r:= (\bar\sigma^2-\sigma^2)/(\bar\sigma^2-1).
$
The conditions in the proposition imply that~$|r|<1$, and hence,~$|r\varepsilon|<\varepsilon$. By~\eqref{eq:w_def}, $w\in\mathrm{span}\left(v^1,v^2\right)$, and this completes the proof.~\hfill{\qed}



\bibliographystyle{IEEEtran}
\bibliography{refs}

\end{document}